\documentclass[epj]{webofc}
\usepackage[varg]{txfonts}   % Web of Conferences font
\usepackage{units}
\begin{document}
%
%\linenumbers
%
\title{(Very)-High-Energy Gamma-Ray Astrophysics: the Future}

\author{Alessandro De Angelis\inst{1}\fnsep\thanks{\email{alessandro.deangelis@pd.infn.it}}}
        % etc.

\institute{INFN Padova, via Marzolo 8, I-35141 Padova (Italy); LIP/IST, Av. Elias Garcia 14, 1000 Lisboa (Portugal)
          }

\abstract{%
Several projects planned or proposed can significantly expand our
knowledge of the high-energy Universe in gamma rays. Construction of the Cherenkov
telescope array CTA is started, and other detectors
are planned which will use the reconstruction of extensive air
showers. This report explores the near future, and possible evolutions   in a longer term.

}
\maketitle
%\tableofcontents

%
\section{Introduction}

Among cosmic rays, photons are particularly important. Being neutral they can travel
long distances without being deflected by galactic and extragalactic magnetic fields,
hence they allow us to directly study their emission sources. The gamma emission is also related to the emission of charged cosmic rays, and to the emission of neutrinos. Finally, gamma-rays can be the signature of new physics at fundamental scales, new particles in particular.  These facts have 
pushed the scientific community during the last years  to study  high-energy gamma rays; many new results have been obtained since the beginning of the millennium, in particular thanks to the imaging atmospheric Cherenkov telescopes (IACTs) H.E.S.S., MAGIC and VERITAS. See \cite{ourbook} for a summary of the main achievements.

Still many problems are however open, in particular:
\begin{itemize}
\item The study of the origin of cosmic rays. Thanks to the first generation of  IACTs and to the AGILE and Fermi satellites, it is well established nowadays that supernova remnants (SNRs) can accelerate cosmic rays up to some 100 TeV. The mechanism of acceleration of particles up to the PeV is still to be established. Above some PeV, it is common belief that the acceleration sites should be extragalactic, but the mechanism, possibly related to the behavior of matter near supermassive black holes (SMBH), is still to be understood. 
\item The
study of the 
 propagation of extragalactic gamma rays, which is a probe of cosmology and of the interaction of gamma rays with the cosmic background. 
 
 Once produced,  photons must travel towards the observer. Electron-positron $(e^-e^+)$ pair production in the interaction of VHE photons off extragalactic background photons is the main source of opacity of the Universe to $\gamma$-rays whenever the corresponding photon mean free path is of the order of the source distance, or smaller.
The cross-section is maximized ($\sigma_{max} \simeq 1.7 \times 10^{-25}$ cm$^2$) for background photons of energy:
 \begin{equation} 
 \label{eq.sez.urto-1}
\epsilon (E) \simeq \left(\frac{900 \, {\rm GeV}}{E} \right) \, {\rm eV}~.
\end{equation}
The maximum photon density corresponds to the CMB; its total density is about 400 photons per cubic centimeter.

A region of particular interest is the so-called extragalactic background light~(EBL), \index{EBL (extragalactic background light)}  i.e., the light in the visible and near infrared regions.
It  consists of the sum of starlight emitted by galaxies throughout their whole cosmic 
history, plus possible additional contributions, like, e.g., light from hypothetical first 
stars that formed before galaxies were assembled.
Therefore, in principle the EBL contains important information about both the evolution of baryonic 
components of galaxies and the structure of the Universe in the pre-galactic era. Note that the  light is redshifted by a factor $(1+z)$ due to the expansion of the Universe, and thus, the  visible light from old sources is detected as infrared.
The density of EBL photons in the region near the visible can be derived by direct deep-field observations, and by constraints on the propagation of VHE photons. A plot of the present knowledge on the density of photons in the EBL region is shown in Fig. \ref{fig:CoppiAharonian}, left. The corresponding opacity of the Universe is illustrated in Fig. \ref{fig:CoppiAharonian}, right. 
%
%Fig. \ref{fig:density}, right, shows a summary of the estimated photon number density of the background photons as composed by the radio background, the cosmic microwave background, and 
%the infrared/optical/ultraviolet background~(EBL).

\begin{figure}
\centering
\includegraphics[width=.36\textwidth]{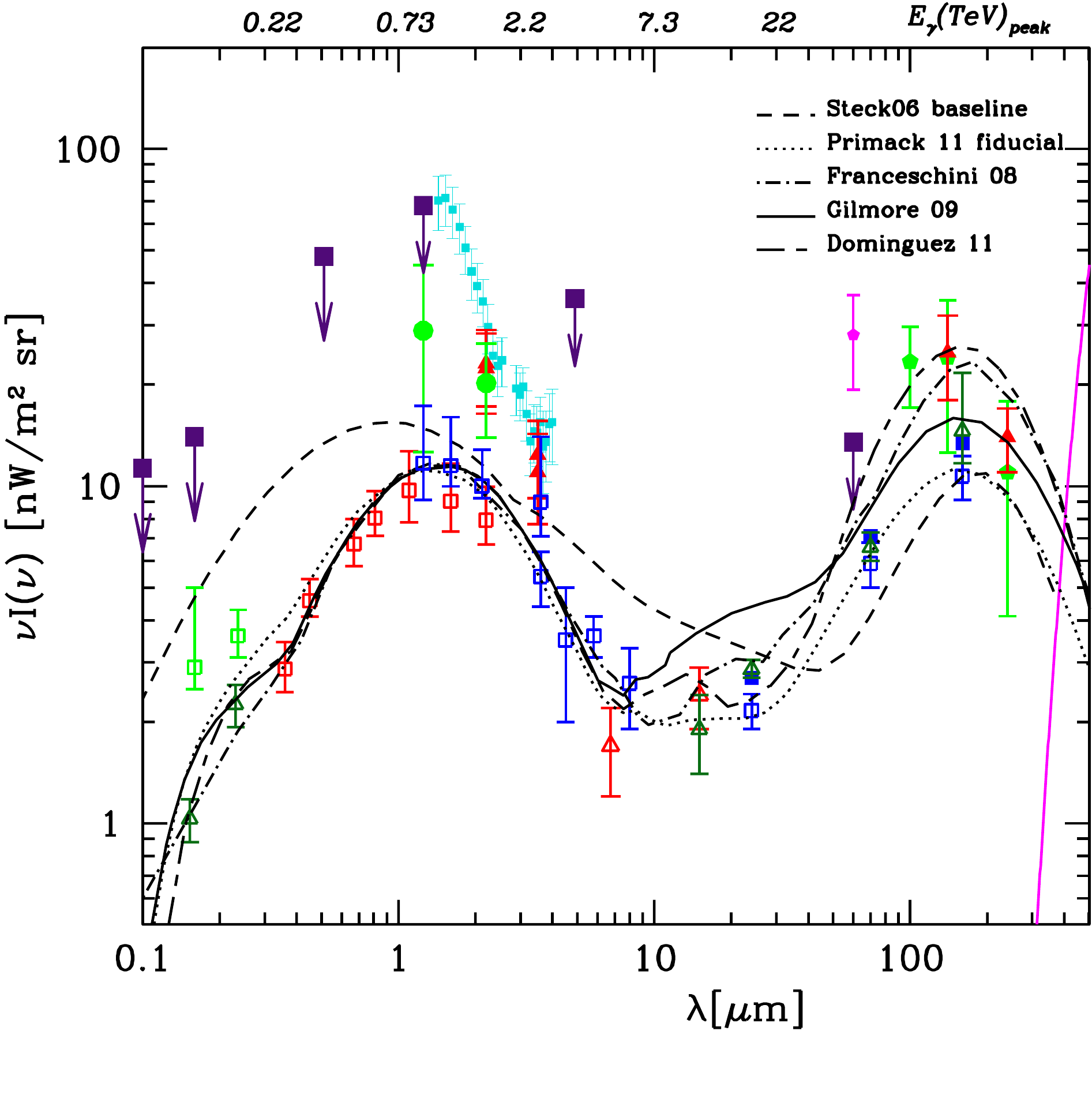}
 \includegraphics[width=.49\textwidth]{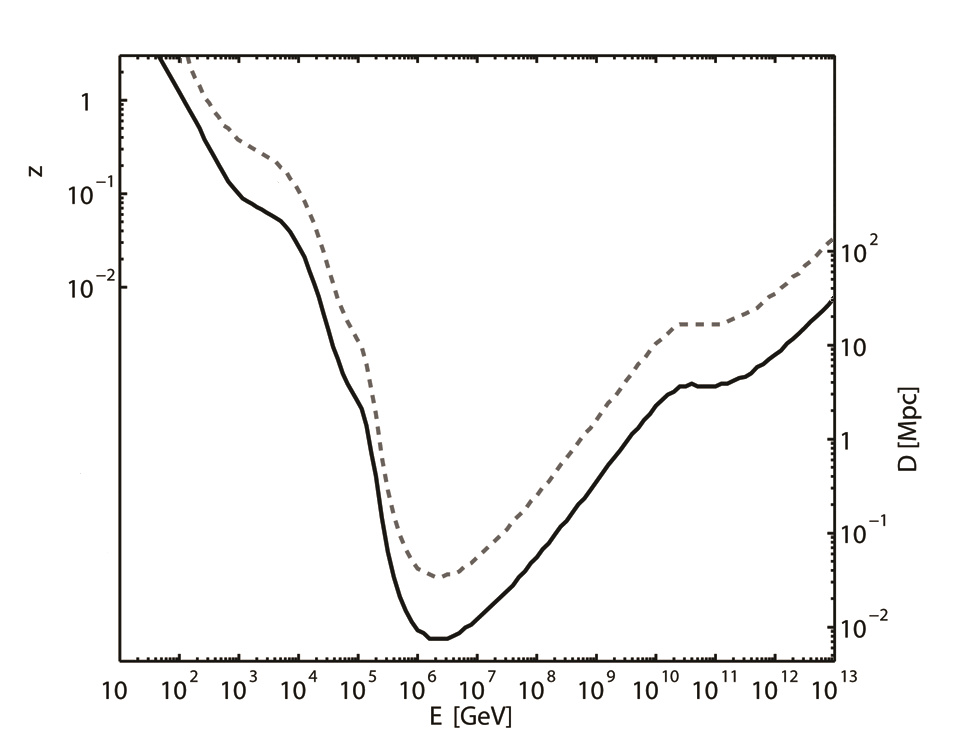}
\caption{\label{fig:CoppiAharonian}
Left:  Spectral energy distribution of the EBL as a function of the wavelength.  Open symbols correspond to lower limits from galaxy counts
while filled symbols correspond to direct estimates.  The curves show a sample of different recent EBL models, as labeled. 
On the upper axis  the TeV energy corresponding to the peak of the $\gamma\gamma$ cross section is plotted. From L. Costamante, IJMPD 22 (2013) 1330025.
 Right:
Curves corresponding to the so-called gamma-ray horizon $\tau(E,z)=1$ (lower) and to a survival probability of $e^{-\tau(E,z)}$ = 1\% (upper). 
Adapted from A. de Angelis, G. Galanti, M. Roncadelli, MNRAS  432 (2013) 3245.}
\end{figure}

% \begin{figure}[tb]
% \centering
% %\includegraphics[height=.44\textwidth]{horizon-05blanch-Magic_Hess}
% %\includegraphics[height=.44\textwidth]{depth-01kneiske_v2_fig3}
% \caption[]%
% {\label{fig:optical-depth}
% Optical depth as a function of the photon energy
% for the source redshifts 0.03, 0.3, 0.5, 1, 2, 3, and 4 (from bottom to top) in the model developed by O.~Blanch and M.~Martinez in the second citation in Ref.~\cite{blanch-vari} (left panel)
% and for a range of source redshifts comprised between 0.03 and 5 (from bottom to top) in the model developed by T.M.~Kneiske and collaborators~\cite{kneiske-proc} (right panel).
% }
% \end{figure}
 \item The examination
of the ultimate nature
of matter and of physics  beyond the Standard Model, in particular  through searches  for dark matter, 
for new particles in general in particular by the determination of the energy density of the vacuum, for the effects of quantum gravity. High-energy gamma astrophysics is sensitive to energy scales important
for particle physics, in particular  the 100 GeV scale expected for
cold dark matter; the TeV scale, where supersymmetric particles
might appear; and finally, it might be possible to access indirectly   the unification scale and the Planck scale: an energy $\sim  10^{19}$ GeV, corresponding to a mass
$\sqrt{\hbar c/G}$. 
\end{itemize}

We shall define in the following as high-energy (HE)  photons the photons above 30 MeV; as very-high-energy (VHE) the photons above 30 GeV.

\begin{figure}
\begin{center}
\includegraphics[width=\textwidth]{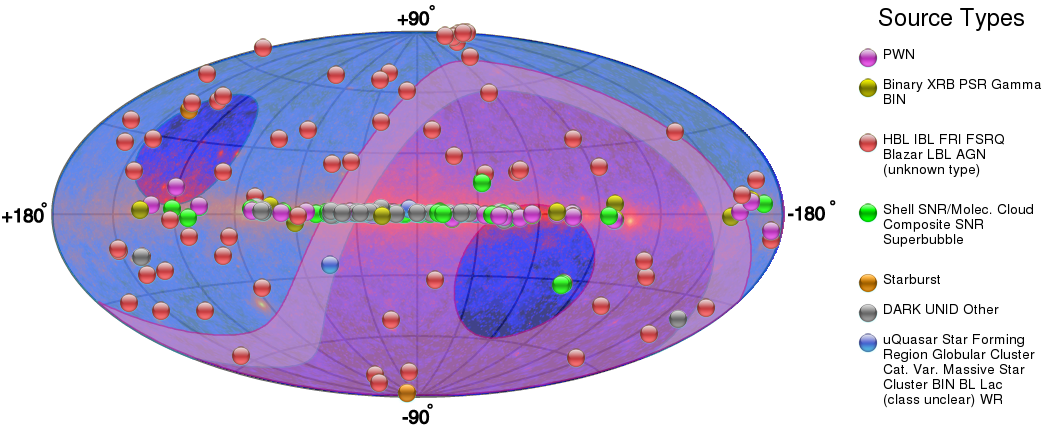}
\end{center}
\caption
{\label{fig:gammamap}Sources of VHE emission plotted in galactic coordinates. 
The background represents the high-energy gamma-rays detected by Fermi-LAT. The clear blue and pink regions correspond respectively to the visible regions within 30$^\circ$ of the zenith from a detector at the tropic in the Northern hemisphere (approximately like MAGIC or VERITAS) and at the Southern hemisphere (approximately like H.E.S.S.).
From http://tevcat.uchicago.edu/, January 2016.}
\end{figure}

A look to the sources of cosmic gamma-rays in the HE region shows a diffuse background, plus a set of localized emitters. Roughly 3000 HE emitters have been identified up to now, mostly by the Fermi-LAT, and some 200 of them are  VHE emitters as well (Fig. \ref{fig:gammamap}).

About half of the VHE gamma-ray emitters are objects in our Galaxy;   most of them can be associated to supernova remnants (SNR).
 The remaining half are extragalactic, and the space resolution of present detectors (slightly better than 0.1$^\circ$) is not good enough to associate them with particular points in the host galaxies; we believe, however, that they are produced in the vicinity of supermassive black holes in the galaxies. The abundance of galactic objects can be explained by the fact that, being closer,  they suffer a smaller attenuation. 

%
%%\subsubsection{Transient phenomena and gamma-ray bursts; quasi-periodical emissions\label{sec:transients}}
%\index{gamma-ray!transients} \index{gamma-ray!flares}
%
%Among cosmic rays, gamma-rays are important not only because they  point to the sources, but also because the sensitivity of present instruments is such that transient events (in jargon, ``transients'') can be recorded. Sources of HE and VHE gamma-rays 
%(some of which might likely be also sources of charged cosmic rays, neutrinos and other radiation) were indeed discovered to exhibit transient phenomena, with timescales from few seconds to few days. % At present, however, only photon transient could be studied, thanks to the comparatively large sensitivity of gamma-ray detectors.
%
The sky exhibits in particular transient events from steady emitters (``flares'') and burst of gamma-rays from previously dark regions (``gamma-ray bursts''). % The phenomenology of such events is described in the rest of this Section.

\section{How to detect photons at different energies:  present and  near future}

Due to the conversion probability in the atmosphere (whose thickness is about 28 radiation length at sea level) only satellite-based detectors can detect primary X/$\gamma$-rays. Satellites are anyway small, being of about 1 m$^2$ their maximum dimensions 
because of the cost of space technology.
 If the energy of the primary particle is large enough, some of
the products of the shower generated by the interaction with the atmosphere, acting as a calorimeter, can reach ground-based detectors.  It is appropriate to build such detectors at high altitudes, where 
atmospheric dimming is lower. The area sampled by ground-based detectors can be much larger than the typical areas of satellites. Since the fluxes of high-energy photons are low and decrease rapidly with increasing energy, 
TeV and PeV gammas can be detected only from the atmospheric showers they produce, i.e., by 
means of ground-based detectors.

%\subsubsection{Night-Sky Background.}

%----------------
\subsection{Satellites}
%----------------

Satellite  gamma telescopes  can detect 
the primary photons at energies lower than ground-based telescopes.

Main figures of merit for a satellite-borne detector are its effective area (i.e., the product of the area 
times the detection efficiency), the energy resolution, the space or angular resolution 
(called as well point-spread function, or PSF).

Detectors on satellites have a small effective area, of order of 1 m$^2$ maximum, which limits the sensitivity.
They have a large duty cycle,  and they 
suffer a low rate of background events, since they can be coupled to anticoincidence systems rejecting the
charged cosmic rays. They have a large cost, dominated by the costs of launch and by the strong requirements of instruments which must be sent in space, with little or no possibility of intervention to fix possible bugs.

The Fermi satellite, in orbit since 2008, has an effective area of  about 1 m$^2$; its function is double, since it can also operate as a trigger for ground-based telescopes.

%----------------
\subsection{Ground-based gamma detectors}
%----------------

Ground-based VHE telescopes such as 
%MILAGRO\index{MILAGRO telescope}, ARGO\index{ARGO telescope}, 
HAWC,
H.E.S.S.\index{H.E.S.S. telescope}, MAGIC\index{MAGIC telescope} and VERITAS\index{VERITAS telescope}, 
detect the atmospheric showers produced by primary photons and cosmic rays of energy higher than 
those observed by satellites.

The two kinds of detectors (on satellite and at ground) are complementary. At energies below 1 GeV or so, the showers generated by photons do not have the time to develop properly, and thus the only way to detect such photons below this energy is related to the use of satellites. Ground-based detectors have a huge effective area, so their sensitivity is high;
they detect a huge amount of background events, but they have low cost.

There are two main classes of ground based VHE gamma detectors: the EAS arrays 
 and the Cherenkov telescopes.

%\begin{figure}
%\centering
%%\includegraphics[width=.7\textwidth]{fig16-eas}
%\caption{\label{fig:EASvsIACT}
%Sketch of the operation of Cherenkov telescopes and of EAS detectors.}
%\end{figure}

%---
\subsubsection{EAS detectors}
%---

The EAS detectors, such as  
HAWC which is presently in operation, 
are large arrays of detectors sensitive to charged secondary particles generated in the atmospheric showers.
They have a high duty cycle and a large field-of-view, but a relatively low sensitivity.
The energy threshold of such detectors is rather large, in the order of a few hundred GeV -- a shower initiated by a 1 TeV photon typically 
has its maximum 8 km a.s.l.

The principle of operation is the same as the one for the UHE cosmic rays detectors like Auger, i.e., 
direct sampling of the charged particles in the shower. This can be achieved:
\begin{itemize} 
\item
either using a sparse array of scintillator-based detectors, Resistive-Plate Chambers (RPC), or water Cherenkov pools;
\item
or by effective covering of the ground, to ensure efficient collection and hence lower the energy threshold.
%\begin{itemize}
%\item
%The ARGO-YBJ detector\index{ARGO-YBJ detector}  at the Tibet site followed this approach.
%It was of an array of resistive plate counters.
%Its energy threshold was in the 0.5 TeV-1 TeV range.
%The Crab Nebula with a significance of about 
%5\,standard deviations ($\sigma$) in 50~days of observation.
%\item MILAGRO  was a water-Cherenkov  instrument  located in New Mexico
%(at an altitude of about 2600~m a.s.l.).
%It detected the Cherenkov light
%produced by the secondary particles of the shower when they enter the water pool instrumented 
%with photomultipliers.
%MILAGRO could detect the Crab Nebula with a significance of about 5\,$\sigma$ in 100~days of 
%observation, at a median energy of about 20~TeV.
%\end{itemize}
\end{itemize}
Possibly a combination of the techniques above can be used, with a central region more dense, to allow discrimination between hadron-generated and photon-generated showers, and a peripheral region.

The energy threshold of EAS detectors is at best in the 0.5 TeV - 1 TeV range, 
so they are built to detect UHE photons as well as the most energetic VHE gammas.
At such energies fluxes are small and large surfaces of order of~10$^4$~m$^2$ are required.

Concerning the discrimination from the charged cosmic ray background,
muon detectors devoted to hadron rejection may be present. Otherwise, it is based on the reconstructed shower shape.
The direction of the detected primary particles is computed from the arrival times with an angular precision of about 1$^\circ$. 
The calibration can be performed by studying the shadow in the reconstructed directions caused by the Moon.
Energy resolution is poor.

Somehow, the past generation EAS detectors were not sensitive enough and just detected a handful of sources. This lesson lead to a new EAS observatory with much larger sensitivity: the High Altitude Water Cherenkov detector HAWC, inaugurated in 2015.
%PA+MP presently under completion.

%The typical energy thresholds of the EAS arrays are rather large

%\paragraph{HAWC.}

\begin{figure} 
\begin{center}
\includegraphics[width=.506\textwidth]{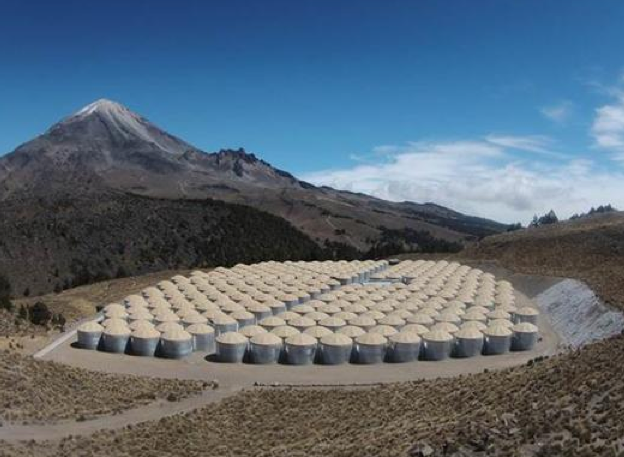}
\includegraphics[width=.37\textwidth]{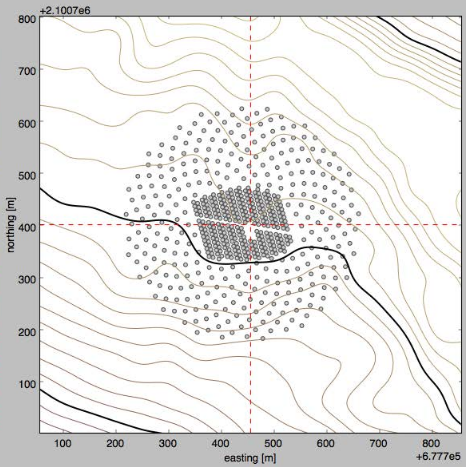}
\end{center}
\caption{\label{fig:hawc}
Left: the HAWC detector. Right: its possible extension.}
\end{figure}

HAWC (Fig. \ref{fig:hawc})\index{HAWC observatory} is
a very high-energy gamma-ray observatory located in Mexico  at about 4100 m asl and latitude $19^{\circ}\,$N . 
It consists of 300 steel tanks of $\unit[7.3]{m}$ diameter and $\unit[4.5]{m}$ deep,
covering an instrumented area of about $\unit[22~000]{m^2}$. Each tank is filled with purified 
water and  contains four 
PMT of 3.8 inch diameter and one high-quantum efficiency 10-inch PMT, which observe the Cherenkov light emitted in water by superluminal particles in
atmospheric air showers.
Photons traveling through the water typically undergo Compton scattering or produce an electron-positron pair,
also resulting in Cherenkov light emission. This is an advantage of the water Cherenkov technique,
as photons constitute a large fraction of the electromagnetic component of an air shower at ground.

HAWC improves the sensitivity for a Crab-like point spectrum by a factor of
15 in comparison to MILAGRO. The sensitivity should be also such to possibly detect gamma-ray burst emissions at high energy.

%A future installation in the Northern hemisphere, a hybrid detector called LHAASO to be deployed in China, is currently under discussion.

%---
\subsubsection{Cherenkov telescopes}\index{Cherenkov!telescope}
%---

Most of the experimental results on VHE photons are presently due to
Imaging Atmospheric Cherenkov Telescopes\index{Imaging Atmospheric Cherenkov Telescope (IACT)} 
(IACTs).

IACTs such as the first successful cosmic detectors, called WHIPPLE\index{WHIPPLE telescope}, and then the second generation instruments 
HEGRA\index{HEGRA telescope} and CANGAROO\index{CANGAROO telescope}, and presently the third generation H.E.S.S.\index{H.E.S.S. telescope}, 
MAGIC\index{MAGIC telescope} and 
VERITAS\index{VERITAS telescope}, detect the Cherenkov photons produced in air by charged, locally superluminal 
particles in atmospheric showers.
For reasons explained below, they have a low duty cycle (about 1000 to 1500 hours per year) and a small field-of-view (FoV), but they have a high 
sensitivity and a low energy threshold.

We refer to \cite{ourbook} for the operation principles and for the performance and the main results of these detectors.

\begin{figure} 
\centering
\includegraphics[width=.9\textwidth]{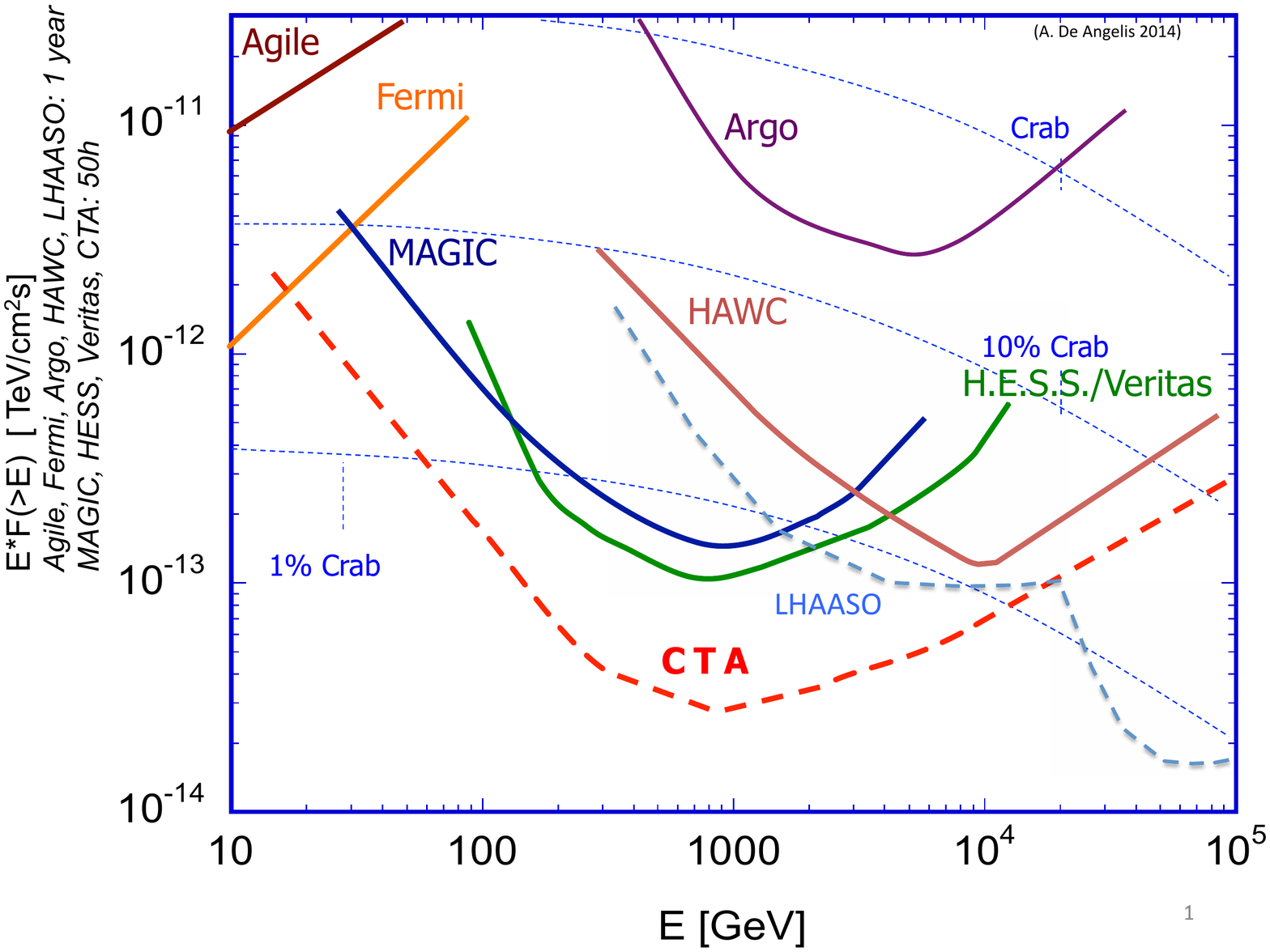}
\caption{\label{fig:Sensitivities}
Sensitivities of some present and future HE gamma detectors, measured as 
the minimum intensity source detectable at 5\,$\sigma$.
The performance for EAS and satellite detector is based on one year of data taking; for Cherenkov 
telescopes it is based on 50 hours of data. From \cite{ourbook}.}
\end{figure}

A simplified comparison of the characteristics of the Fermi LAT satellite detector, of the 
IACTs and of the EAS detectors (ground-based), is given in Table \ref{tab:DetectorsComparison}.
The sensitivities of the above described high-energy detectors are shown in 
Fig.  \ref{fig:Sensitivities}.

\begin{table}
\begin{center}
\begin{tabular}{ | l | l | l | l | }
                \hline &&& \\[-.9em]
{\bf Quantity}        & {\bf Fermi}       & {\bf IACTs}          & {\bf EAS}
\\ \hline  &&& \\[-.9em]
Energy range          & 20 MeV--200 GeV & 100 GeV--50 TeV    & 400 GeV--100 TeV
\\         &&& \\[-.9em]
Energy res.     & 5-10\%            & 15-20\%        & $\sim$ 50\%
\\         &&& \\[-.9em]
Duty Cycle            & 80\%              & 15\%                 & $>$ 90\%
\\         &&& \\[-.9em]
FoV                   & $4 \pi / 5$       & 5$^\circ$ $\times$ 5$^\circ$ & $4 \pi / 6$
\\         &&& \\[-.9em]
PSF       & 0.1$^\circ$           & 0.07$^\circ$             & 0.5$^\circ$
\\         &&& \\[-.9em]
Sensitivity  & 1\% Crab (1 GeV)  & 1\% Crab (0.5 TeV)   & 0.5 Crab (5 TeV)
\\        \hline
\end{tabular}
\end{center}
\caption{\label{tab:DetectorsComparison}
A comparison of the characteristics of Fermi, the IACTs and of the EAS particle detector 
arrays. Sensitivity computed over one year for Fermi and the EAS, and over 50 hours for the IACTs. 
}
%\begin{tabular}{ @{} c @{\,} l @{} c}
%($^*$)     & Decreases to 15\% after cross-calibration with GLAST \cite{noimagglast}.                          & \hfil
%\\
\end{table}

The construction of the next generation IACT, called the Cherenkov Telescope Array (CTA), has already started.

\subsection{The near future of VHE astrophysics with Cherenkov telescopes: CTA}

 The CTA \cite{CTA} is a future instrument for VHE
  gamma astrophysics that is expected to provide
  an order of magnitude improvement in sensitivity over existing instruments (Fig. \ref{fig:Sensitivities}).

CTA is a robust collaboration, including more than 1000 scientists from all around the world, and large part of its funding (above 200 MEUR)  is granted. 
CTA is included in the 2008 roadmap of the European Strategy Forum on Research Infrastructures (ESFRI). It is one of the ``Magnificent Seven''
experiments  of the European strategy for astroparticle physics published by ASPERA, and highly ranked in the strategic plan for European astronomy of ASTRONET. In addition CTA is a recommended project for the next decade in the US National Academies of Sciences Decadal Review.

  An array of tens of telescopes will detect gamma-ray induced showers over a large area on the ground, 
increasing the efficiency and the sensitivity, while providing a much larger number of views of each cascade. 
This will result in both improved angular resolution and better suppression of charged cosmic-ray background events.
 
In the present design scenario, CTA will be deployed in two sites. 
The southern hemisphere array, possibly in Paranal, will consist of three types of telescopes with different mirror sizes, in order to cover 
the full energy range. In the northern hemisphere array, possibly in the Canary island of La Palma, the two larger telescope types would be deployed.
 \begin{itemize}
\item The low energy (the goal is to detect showers starting from an energy of 20 GeV)  instrumentation will consist of  four 23 meter telescopes with a  FoV of about 4-5 degrees.
\item The medium energy range, from around 100 GeV to 1 TeV, will be covered by some 15-25 telescopes of the 12 meter class with a FoV of 6-8 degrees. An alternative design uses smaller reflectors (abot 9 m) with a secondary optics (Schwarzschild-Couder design, \cite{vladi}).
\item The high energy instruments, operating above 10 TeV, will consist of a large number (of the order of 50) of small (4-6 meter diameter) telescopes with a FoV of around 10 degrees. Also in this case single-reflector designs are in competition with Schwarzschild-Couder design.
\end{itemize}
Prototyping is advanced for all these telescopes, but there is still room for technical improvement (SiPM in particular).

The southern CTA will cover about three square kilometers of land with around 60 telescopes that will monitor all the energy ranges in the center of the Milky Way's galactic plane. The northern CTA will cover one square kilometer and be composed of some 20 telescopes. These telescopes will be mostly targeted at extragalactic astronomy.

\begin{figure} 
\begin{center}
\includegraphics[width=.60\textwidth]{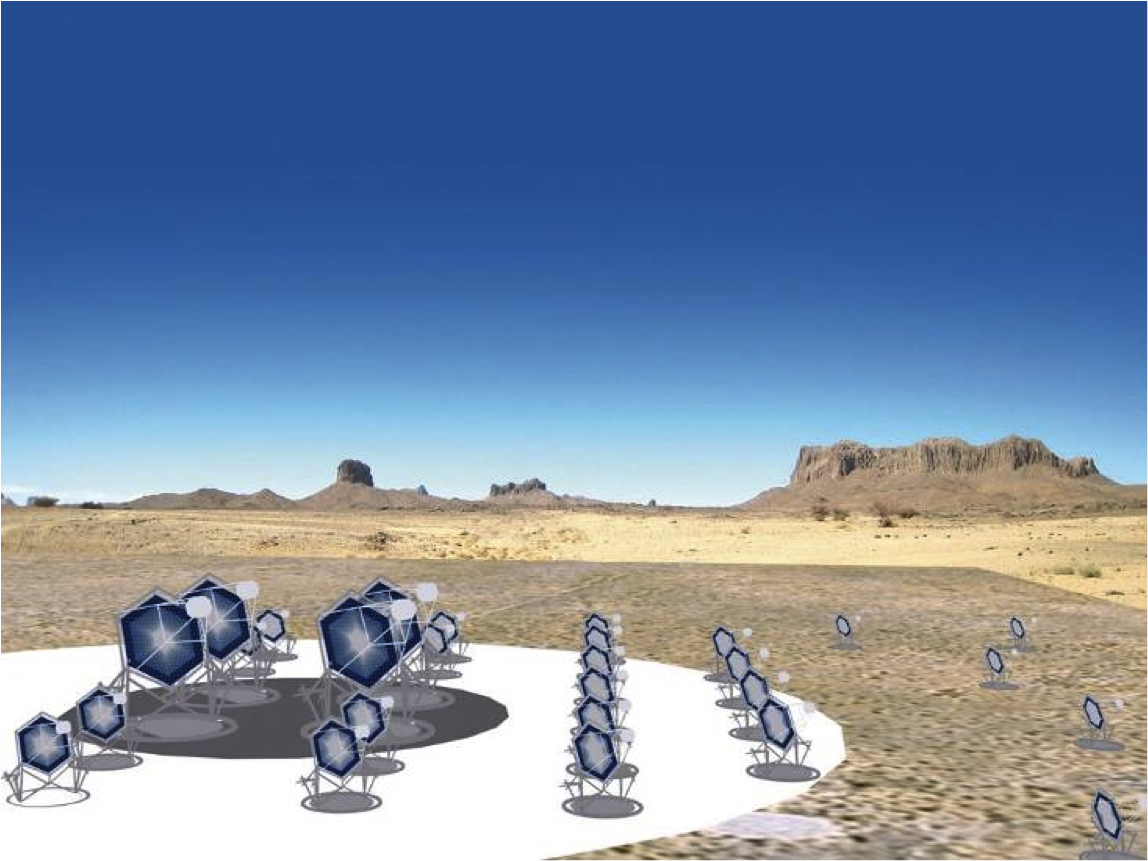}
\includegraphics[width=.37\textwidth]{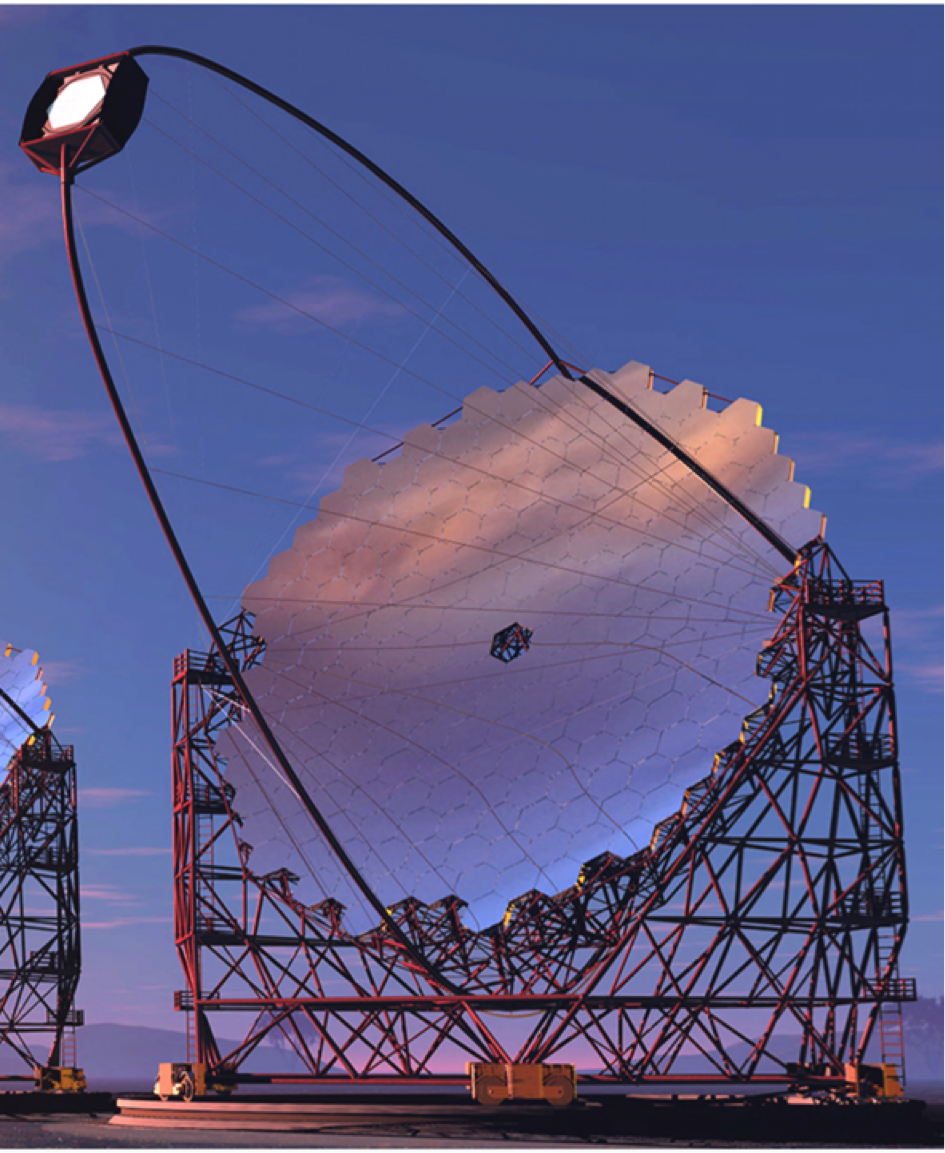}
\end{center}
\caption{\label{fig:cta}
Left: possible layout of the CTA. Right: project of the large-size telescope (LST). Credit: CTA Collaboration.}
\end{figure}

The telescopes of different sizes will be disposed in concentrical circles, the largest in the center (Fig. \ref{fig:cta}). Different modes of operation will be possible: deep field observation; pointing mode; scanning mode.

CTA is expected to surpass in performance the present IACTs around 2020.

\subsection{The near future of EAS detectors} %%%%%%%%%%%%%%%%%%%%%%%%

There is a lot of activity on the development of EAS detector, since they are relatively cheap (of the order of 30 MEUR) and we know now what is the correct size for the detection of a relevant number of sources.

\subsubsection{HAWC+}

An upgrade of the HAWC high-energy gamma-ray observatory with a sparse array of small
outrigger tanks is being investigated \cite{hawcp}. For showers where the core falls outside the array there are
ambiguities in the reconstruction between the core position, the shower angle and the shower size
or energy. An outrigger array can determine the core position for showers falling outside the main
array elevating the ambiguities and making these showers well reconstructable.  A gain of 3-4 in
sensitivity for gammas above 10 TeV can be obtained over what is presently achieved, and such a detector can be built in 2-3 years. Funds are already available.

\subsubsection{LHAASO}

LHAASO (Large High Altitude Air Shower Observatory) is planned to  be located at about 4400 m asl and latitude $30^{\circ}\,$N in the Daochen site,  China.
It will consist of the following major components \cite{lhaaso,pino}:
\begin{itemize}
\item 1 km$^2$ array (LHAASO-KM2A), including 5600 scintillator detectors, with 15 m spacing, for electromagnetic particle detection.
\item An overlapping 1 km$^2$ array of 1200, 36 m$^2$ underground water Cherenkov tanks, with 30 m spacing, for muon detection (total sensitive area 40,000 m$^2$).
\item A close-packed, surface water Cherenkov detector facility with a total area of 90,000 m$^2$ (LHAASO-WCDA), four times that of HAWC.
\item 24 wide field-of-view air Cherenkov (and fluorescence) telescopes (LHAASO-WFCTA).
\item 452 close-packed burst detectors, located near the center of the array, for detection of high energy secondary particles in the shower core region (LHAASO-SCDA).
\end{itemize}
%LHAASO The start of data taking is expected 2 -- 3 years after the start of installation planned at the beginning of 2016. The completion of installation in 6 years.

%The sensitivity of LHAASO to point--like gamma ray sources is shown in Fig. \ref{fig:lhaaso_sens} where is compared to other experiments \cite{cui14}. The sensitivity curve has been calculated for a Crab Nebula like energy spectrum (power law with exponent $-2.63$) extending to PeVs without any cutoff. 

A configuration corresponding to 25\% of the final detector might be ready by 2019 - 2020. It is crucial that the completion of the first 25\% of the detector (nominally foreseen for 2018) is not too much delayed with respect to the extension of HAWC, otherwise most of its scientific impact might be scooped.

\subsubsection{HiSCORE}

HiSCORE (Hundred*i Square-km Cosmic ORigin Explorer),
located in the Tunka Valley near Lake Baikal at about 3200 m asl and latitude $51^{\circ}\,$N,  will consist of an array of wide-angle ($\Omega\sim$ 0.6-0.85 sr) light-sensitive detector stations, distributed over an area of the order of 100 km$^2$. A HiSCORE detector station consists of four photomultiplier tubes, each equipped with a light-collecting Winston cone of 30$^{\circ}$ half-opening angle  \cite{hiscore}. 

The primary goal of this non-imaging Cherenkov detector is gamma-ray astronomy in the 10 TeV to several PeV range. A prototype array of 9 wide-angle optical stations, spread on a 300$\times$300 m$^2$ area, has been deployed since October 2013, and technical tests are underway. 
A 1 km$^2$ engineering array has been deployed and is in commissioning.

\begin{figure}
  \centerline{\includegraphics[width=0.5\textwidth]{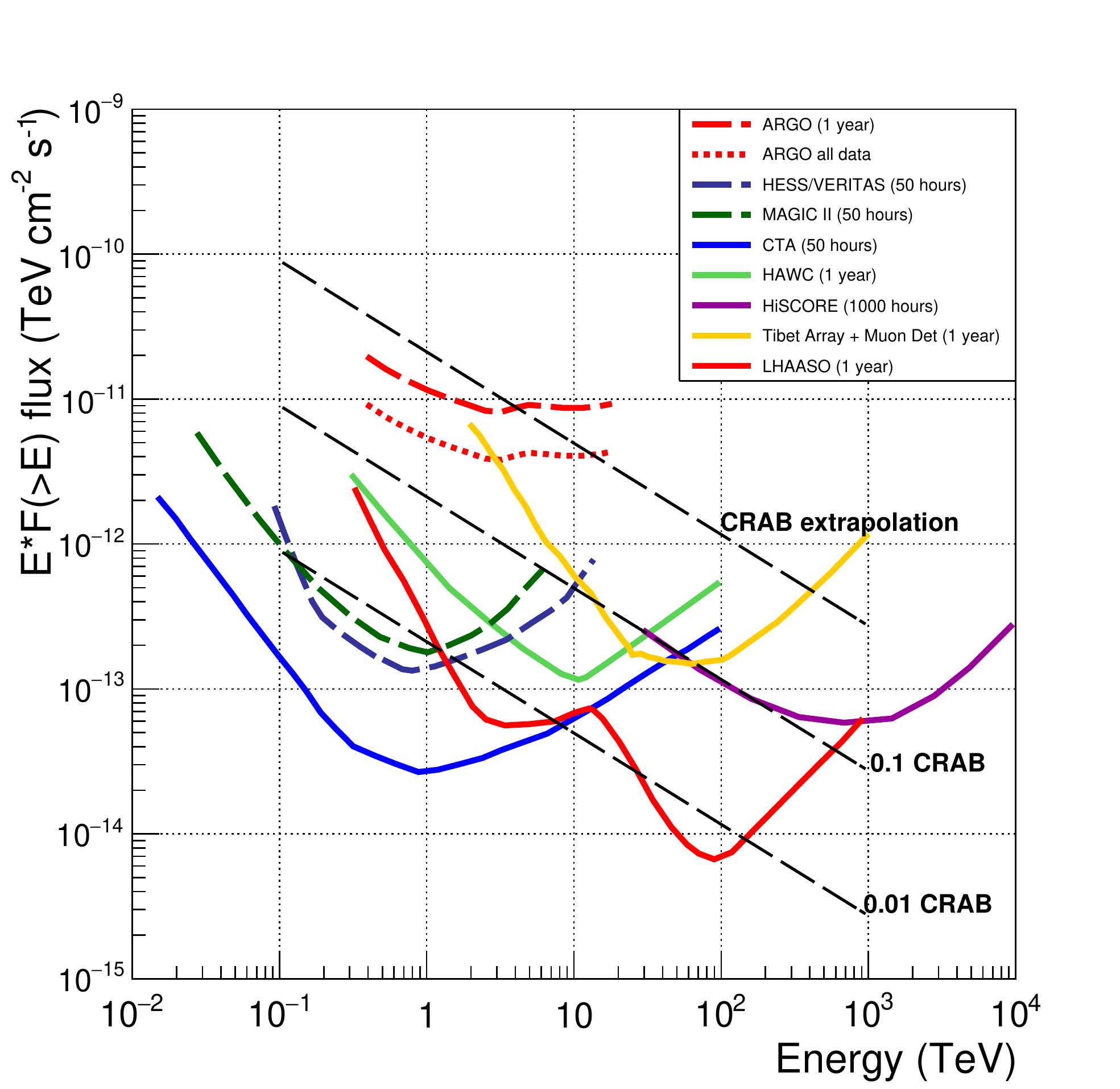}}
    \caption{Sensitivity of ground-based $\gamma$-ray detectors to a Crab-like point source, not considering the attenuation 
    of the source at large energies (from \cite{pino}).}
\label{fig:lhaaso_sens}
\end{figure}

The sensitivity of the HiSCORE standard configuration to point-sources is shown in Fig. \ref{fig:lhaaso_sens}. The curve has been calculated for 1000 h of exposure time, corresponding to  roughly 1.4 years of serendipitous mode operation (see \cite{hiscore} for details).

HiSCORE will be possibly  incorporated in TAIGA (Tunka Advanced Instrument for cosmic ray physics and Gamma Astronomy), a  hybrid detector system
including  in particular IACTs for sensitivity to photons of lower energies.

\subsubsection{Possible EAS experiments in the Southern hemisphere}

Several proposals are being formulated now for a EAS detector at South, which would be the key to study the VHE emission from the Galaxy. HAWC has demonstrated that present EAS technology can be competitive with the technology of large-area coverage by small Cherenkov telescopes.

In particular among the proposals are:
\begin{itemize}
\item a Southern site for HAWC, HAWC-South, to be located in the Atacama desert \cite{hawcs};
\item LATTES, a hybrid water Cherenkov-RPC detector to be located in Southern America \cite{lattes}.
\end{itemize}

\section{A view in a longer term}

Let us analyze the possible future of gamma-ray astrophysics in, say, 15 years from now.

%We shall analyze the different energy ranges in particular with respect to the future of gamma-ray astrophysics in the following subsections.
Gamma rays can be divided in  5 different energy regions, subjectively
chosen
from
the
requirements
of
detection
techniques
and of main scientific issues.
These energy regions are:
%\begin{enumerate}
(a) {Till a few ($\sim$30) MeV;}
(b) {GeV: from 30 MeV to 30 GeV:}
(c) {Sub-TeV: from 30 GeV to 300 GeV;}
(d) {TeV: from 300 GeV to 30 TeV;}
(e) {PeV: from 30 TeV to a few PeV.}
%\end{enumerate}

\subsection{The region till a few MeV}

This region has important implications on the science at the TeV and above, since a good knowledge of the spectra in the MeV
region can constrain the fit to the emitted spectra at high energies, thus allowing:
\begin{itemize}
\item to evidence additional contributions from new physics (dark matter in particular);
\item to estimate cosmological absorption (due to EBL or to possible interactions with axion-like fields \cite{darma}).
\end{itemize}

On top of this, The 0.3-30 MeV energy range is important per se, but experimentally difficult to study. It requires an efficient instrument working in the Compton regime with an excellent background subtraction, and possibly with sensitivity to the measurement of polarization.
Since COMPTEL, which operated two decades ago, no space instrument obtained extra-solar gamma-ray data in the few MeV range; now
we are able to build an instrument one-two orders of magnitudes more sensitive than COMTEL based on Silicon technology, state-of-the-art analog readout, and efficient data acquisition. 

Several proposals (ASTROGAM \cite{astrogam}, COMPAIR \cite{compair}, ...) have been made, and convergence is likely for an experiment to be launched around 2025.

%Nuclear astrophysics and gamma ray spectroscopy; photometry
\subsection{The GeV region}

It is difficult to think for this century of an instrument for GeV photons improving substantially the performance of the Fermi LAT: the cost of space missions is such that the size of Fermi cannot be reasonably overcome with present technologies. New satellites in construction (like the Russian-Italian mission 
GAMMA400 \cite{gamma400} and the Chinese mission HERD \cite{herd}\index{DAMPE mission})  improve some of the aspects of Fermi, e.g., calorimetry. For sure a satellite in the GeV region with sensitivity comparable with Fermi will be needed.

\subsection{The sub-TeV and TeV regions}

CTA appears to have no rivals for the gamma astrophysics in the 
Sub-TeV  and TeV (from a few GeV to a few TeV) energy regions. These are crucial regions for fundamental physics, and for astronomy; they
include also the region where sensitivity is maximum.

CTA will be probably upgraded including state-of-the art photon detection devices of higher efficiencies with respect to the present ones.

There is some room for new concepts of Cherenkov telescopes, in particular:
\begin{itemize}
\item a possible array of 4 or 5 very large detectors at 5000 m in order to boost the sensitivity around in the sub-TeV region (in particular from 20 GeV to 100 GeV) \cite{fiveatfive}. This would boost the science potential in particular in sectors related to fundamental physics (dark matter, photon propagation from far sources);
\item a large-FoV detector to guarantee good coverage of the sky in the TeV region \cite{machete}.
\end{itemize}

\subsection{The PeV region}
 Due to the opacity of the Universe to gamma rays, less than a handful of  sources could be visible in the Northern sky, and less than a dozen in the Southern sky, all galactic. The experiments in the Northern hemisphere (HAWC with its likely extension to HAWC+, LHAASO, TAIGA/HiSCORE if the extension up to 10 km$^2$ will be funded) provide an appropriate coverage of the Northern sky. 
 
The situation in the Southern hemisphere  has room for improvement.   An EAS detector in the South might give substantial input with respect to the knowledge of the gamma sky, and of possible Pevatrons in the Galaxy, and outperform in this sense the SSTs of CTA. Several proposals are being formulated now, and they should join. A large detector in Southern America could compete in sensitivity with the small-size telescopes of CTA-South already at 100 TeV, offering in addition a serendipitous approach.

%\section*{Acknowledgements}

 %\bibliography{biblio.bib}
%
%%%%%%%%% Non-BibTeX users please use
%%%%%%%%%
%%%%%%%%\begin{thebibliography}{}
%%%%%%%%%
%%%%%%%%% and use \bibitem to create references.
%%%%%%%%%
%%%%%%%%\bibitem{RefJ}
%%%%%%%%% Format for Journal Reference
%%%%%%%%Journal Author, Journal \textbf{Volume}, page numbers (year)
%%%%%%%%% Format for books
%%%%%%%%\bibitem{RefB}
%%%%%%%%Book Author, \textit{Book title} (Publisher, place, year) page numbers
%%%%%%%%% etc
%%%%%%%%\end{thebibliography}

\end{document}